\documentclass[prl,
reprint,
 amsmath,amssymb,
 aps,
]{revtex4-2}

\usepackage{graphicx}% Include figure files
\usepackage{dcolumn}% Align table columns on decimal point
\usepackage{bm}% bold math
\usepackage{physics}
\usepackage[colorlinks=true, citecolor=red, linkcolor=blue, urlcolor=black]{hyperref}
\usepackage{here}

\begin{document}

%\preprint{APS/123-QED}

\title{Instability of the Haldane Phase: Roles of Charge Fluctuations and Hund’s Coupling}

\author{Satoshi Nishimoto}
\email{s.nishimoto@ifw-dresden.de}
\affiliation{Department of Physics, Technical University Dresden, 01069 Dresden, Germany}
\affiliation{Institute for Theoretical Solid State Physics, IFW Dresden, 01171 Dresden, Germany}

\date{\today}% It is always \today, today,

\begin{abstract}
We systematically investigate the stability of the symmetry-protected topological (SPT) Haldane phase in spin-1/2 Heisenberg and half-filled Hubbard ladders coupled by ferromagnetic Hund's interactions. Using density-matrix renormalization group (DMRG) method, we analyze key signatures of the Haldane phase: long-range string order, finite spin gap, and characteristic entanglement spectrum degeneracies. In spin-only Heisenberg ladders, we find immediate onset and continuous strengthening of the Haldane phase with increasing Hund's coupling. In contrast, the inclusion of charge fluctuations in Hubbard ladders leads to a nontrivial stability regime, revealing a robust yet bounded region where SPT order persists despite significant charge fluctuations. We identify distinct boundaries separating a trivial insulating phase from the Haldane SPT phase, governed by both Coulomb repulsion and Hund's coupling. Our results highlight the subtle interplay of spin and charge degrees of freedom in correlated itinerant systems and establish essential criteria for observing Haldane physics experimentally in fermionic ladder materials.
\end{abstract}

\maketitle

%\tableofcontents

{\it Introduction.}---
Understanding the robustness of topological phases in the presence of competing interactions and additional degrees of freedom remains a central challenge in condensed matter physics. Symmetry-protected topological (SPT) phases, exemplified by the Haldane phase~\cite{Haldane1983-1,Haldane1983-2} in spin-1 chains, exhibit hallmark features including a finite spin gap~\cite{Affleck1987,White1993}, nonlocal string order~\cite{den_Nijs1989,Kennedy1992,Oshikawa1992}, and fractionalized edge states~\cite{Kennedy1990,Affleck1989} protected by discrete symmetries~\cite{Pollmann2010,Schuch2011,Chen2011,Pollmann2012}. While these properties have been thoroughly established for idealized spin-only models, their stability in realistic materials featuring significant charge fluctuations and orbital degrees of freedom remains poorly understood and is a highly nontrivial issue~\cite{Zirnbauer2021}.

Initially, experimental realization of the Haldane phase occurred predominantly in transition-metal oxides~\cite{Renard1987,Zaliznyak2001,Kenzelmann2002,Kenzelmann2003,Nag2022}. More recently, synthetic quantum systems such as metal-organic frameworks~\cite{Tin2023}, organic molecular chains~\cite{Zhao2023}, and ultracold atomic ladders~\cite{Sompet2022} have demonstrated essential characteristics of the Haldane phase, thereby prompting critical new questions regarding its robustness under reduced electron correlations and enhanced charge fluctuations. Resolving the stability and topological integrity of the Haldane phase in these realistic contexts is thus pivotal both for fundamental insights into quantum magnetism and for guiding the development of novel topological quantum materials.

Recent theoretical investigations have focused intensively on the robustness of SPT phases in specialized fermionic ladder systems, especially regarding the Haldane phase~\cite{Onishi2004,Anfuso2007,Moudgalya2015,Jazdzewska2023}. These studies underscore the crucial roles played by orbital degeneracy, charge dynamics, and symmetry protection. Nevertheless, a comprehensive, quantitative understanding of how charge fluctuations, governed specifically by Coulomb repulsion and Hund's coupling, affect the stability or potential breakdown of the Haldane phase is still lacking.

In this Letter, we systematically investigate the stability of the SPT Haldane phase by performing density-matrix renormalization group (DMRG)~\cite{dmrg} calculations on two complementary ladder models: a Heisenberg ladder to clarify the role of Hund’s coupling, and a Hubbard ladder to elucidate the effects of charge fluctuations as well as Hund’s coupling. We accurately determine the phase boundaries separating the SPT Haldane phase from a trivial Mott/Hund insulating phase, explicitly demonstrating the breakdown of the Haldane phase under sufficiently strong charge fluctuations. Clarifying these conditions is crucial not only for a deeper understanding of correlated topological phases but also as foundational groundwork for exploring unconventional superconductivity and spin-singlet pairing in doped Haldane systems~\cite{Frahm1998,Ammon2000,Payen2000,Patel2020,Zhang2022,Laurell2024}.

\begin{figure}[b]
	\centering
	\includegraphics[width=0.8\linewidth]{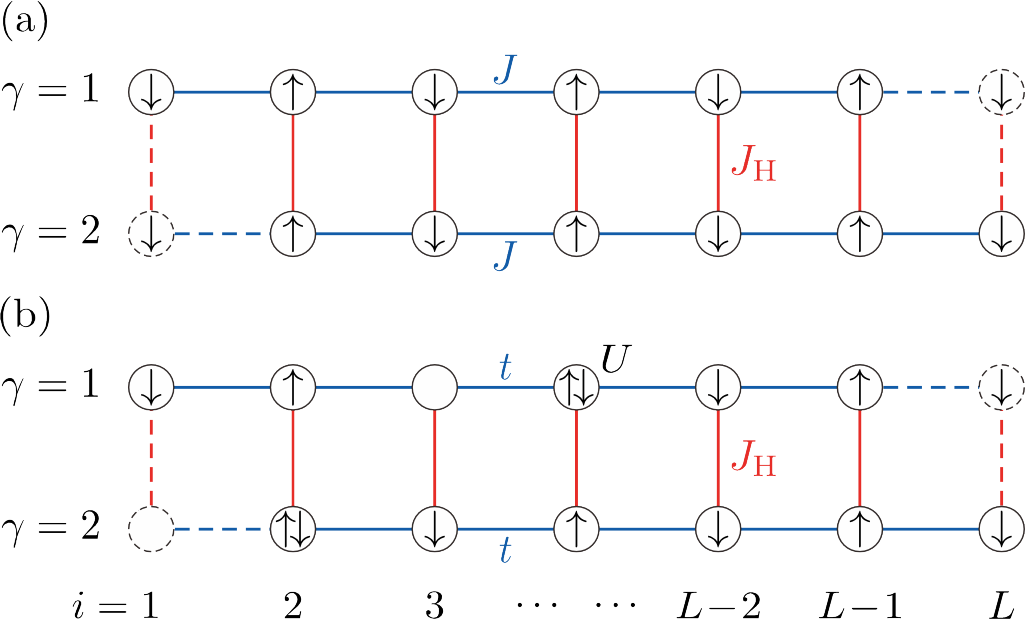}
	\caption{
		Lattice structures of (a) the two-leg Heisenberg ladder and (b) the two-leg Hubbard ladder, with FM Hund’s coupling $J_H$ connecting the chains. Indices $\gamma$ and $i$ label the leg and site positions, respectively. Removing the dashed bonds corresponds to replacing the edge spins with spin-1/2 degrees of freedom in the effective spin-1 Heisenberg chain representation.
	}
	\label{fig:lattice}
\end{figure}

{\it Model definitions and effective mapping.}---
We consider two models of coupled one-dimensional (1D) chains with interchain FM Hund’s coupling: a spin-$\frac{1}{2}$ Heisenberg ladder and a half-filled Hubbard ladder. The Heisenberg ladder serves as a spin-only reference system, while the latter incorporates charge degrees of freedom and their fluctuations.

The Heisenberg ladder is described by
\begin{align}
H_{\rm Heis} = J \sum_{\gamma=1,2} \sum_{i} \mathbf{S}_{\gamma,i} \cdot \mathbf{S}_{\gamma,i+1} - J_{\rm H} \sum_i \mathbf{S}_{1,i} \cdot \mathbf{S}_{2,i},
\label{eq:HamHeis}
\end{align}
where $\mathbf{S}_{\gamma,i}$ denotes a spin-$\frac{1}{2}$ operator on leg $\gamma$ and site $i$, $J>0$ is the antiferromagnetic (AFM) intrachain coupling, and $J_{\rm H}>0$ is the FM Hund's coupling along the rungs. The corresponding Hubbard ladder is given by
\begin{align}
\nonumber
H_{\mathrm{Hub}} = -t \sum_{\gamma=1,2} \sum_{i,\sigma} \left( c^\dagger_{\gamma,i,\sigma} c_{\gamma,i+1,\sigma} + \text{H.c.} \right)\\
 + U \sum_{\gamma,i} n_{\gamma,i,\uparrow} n_{\gamma,i,\downarrow}- J_{\rm H} \sum_i \mathbf{S}_{1,i} \cdot \mathbf{S}_{2,i},
 \label{eq:HamHub}
\end{align}
where $c^\dagger_{\gamma,i,\sigma}$ ($c_{\gamma,i,\sigma}$) creates (annihilates) a fermion with spin $\sigma$ on leg $\gamma$ and site $i$, $t$ is the intrachain hopping amplitude, and $U$ is the on-site Coulomb repulsion. Schematic illustrations of the ladder geometry are shown in Fig.~\ref{fig:lattice}.

In the Heisenberg ladder \eqref{eq:HamHeis}, the FM Hund’s coupling favors the formation of rung triplets. The low-energy Hilbert space is then approximately restricted to the spin-triplet subspace on each rung~\cite{Strong1992}: $\ket{1}_{i} = \ket{\uparrow}_{1,i} \ket{\uparrow}_{2,i}$, $\ket{0}_{i} = (\ket{\uparrow}_{1,i} \ket{\downarrow}_{2,i} + \ket{\downarrow}_{1,i} \ket{\uparrow}_{2,i})/\sqrt{2}$, and $\ket{-1}_{i} = \ket{\downarrow}_{1,i} \ket{\downarrow}_{2,i}$. Projecting onto this subspace yields an effective spin-1 Heisenberg chain:
\begin{align}
	H_{\rm eff} = \frac{J}{2} \sum_i \tilde{\mathbf{S}}_i \cdot \tilde{\mathbf{S}}_{i+1} - \frac{J_{\rm H}}{4}L,
	\label{eq:s1hcain}
\end{align}
where $\tilde{\mathbf{S}}_i = \mathbf{S}_{1,i} + \mathbf{S}_{2,i}$ is a spin-1 operator on rung $i$, and $L$ is the number of rungs. This emergent spin-1 mechanism differs from materials with intrinsic local spin-1 ions (e.g., Ni$^{2+}$), yet at low energies the two are adiabatically connected and realize the same Haldane SPT phase, as elaborated in the Supplemental Material~\cite{SM}.

A similar mapping is expected for the Hubbard ladder at large $U/t$ and/or $J_{\rm H}/t$. In this regime, charge fluctuations are suppressed: the on-site repulsion $U$ penalizes double occupancy, while FM Hund’s coupling $J_{\rm H}$ favors rung spin-triplet formation, stabilizing effective spin-1 moments. Although the precise coupling range for the validity of this spin-only mapping is not known, complete suppression of charge fluctuations is not necessary; dominance of rung triplet states suffices. This provides a controlled platform to explore how residual charge fluctuations impact SPT order.

{\it Numerical methods and observables.}---
We perform extensive DMRG simulations on open ladders with up to $L \times 2 = 241 \times 2$ sites, keeping up to $\chi=10000$ density-matrix eigenstates. 
To characterize the Haldane phase, we calculate four key observables: the string correlation function, spin gap, entanglement spectrum (ES), and spin-spin correlation functions. Numerical values are extrapolated carefully to $\chi \to \infty$ where necessary.

The string correlation function is defined as
\begin{align}
\mathcal{O}_\mathrm{str}(i,j)= \left\langle \tilde{\mathbf{S}}^z_i \exp\left( i\pi \sum_{k=i+1}^{j-1} \tilde{\mathbf{S}}^z_k \right) \tilde{\mathbf{S}}^z_j \right\rangle.
\end{align}
whose asymptotic value $|\mathcal{O}_\mathrm{str}(-\infty,\infty)|$ serves as the string order parameter. In practice, when the order is robust, $|\mathcal{O}_\mathrm{str}(i,j)|$ rapidly reaches a flat plateau as $|i-j|$ increases on large open ladders; in this case we read off the plateau value from a fixed, sufficiently long ladder and take it as the order parameter. When the plateau is not unambiguous (e.g., near phase boundaries), we determine the order parameter by finite-size scaling of a central estimator evaluated on $L\times2$ ladders (see the Supplemental Material for details~\cite{SM}). For the spin-1 Heisenberg chain, this parameter takes the value $|\mathcal{O}_\mathrm{str}^{S=1}(-\infty,\infty)| \approx 0.3743$~\cite{White1993,Kolezhuk2002}.

The spin gap is obtained from the singlet–triplet excitation energies, defined as $\Delta_{\rm s}=\lim_{L\to\infty}[E_0(L,1)-E_0(L,0)]$, where $E_0(L,S^z)$ is the ground-state energy of a ladder with $L$ rungs and total spin $S^z$ along the $z$-axis. To eliminate zero-energy excitations associated with spin-$\frac{1}{2}$ edge states, we remove one spin-$\frac{1}{2}$ site from each ladder edge, leaving effective spin-$\frac{1}{2}$ degrees of freedom on the outermost rungs [see Fig.~\ref{fig:lattice}]. The spin gap for the spin-1 Heisenberg chain [Eq.~\eqref{eq:s1hcain}] is well-established as $\Delta_{\rm s}^{S=1}/J=0.20523962$~\cite{White1993,Ueda2011,Ejima2015}.

Additionally, we examine the ES~\cite{Li2008}, obtained from the Schmidt decomposition of the ground state upon bipartitioning: $|\Psi\rangle = \sum_{\alpha} \lambda_\alpha |\phi_\alpha^{\rm L}\rangle \otimes |\phi_\alpha^{\rm R}\rangle$. The eigenvalues $\lambda_\alpha^2$ of the reduced density matrix $\rho_L=\mathrm{Tr}_R |\Psi\rangle\langle\Psi|$ define the ES. In the Haldane phase, the ES displays characteristic even-fold degeneracies, directly indicating fractionalized spin-$\frac{1}{2}$ edge states and confirming SPT order.

\begin{figure}[t]
	\centering
	\includegraphics[width=0.8\linewidth]{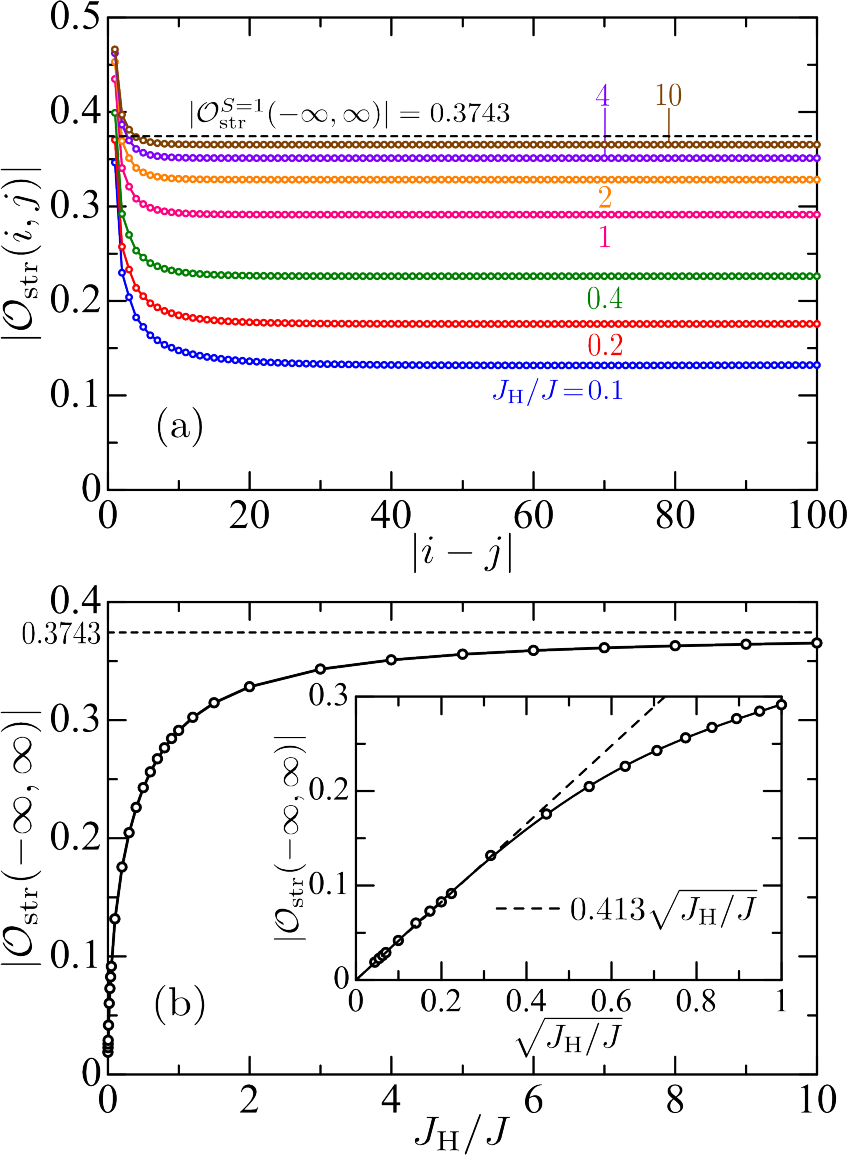}
	\caption{
		String order in the Heisenberg ladder.
		(a) String correlation function $\mathcal{O}_\mathrm{str}(|i-j|)$ plotted as a function of distance $|i-j|$ for several values of $J_{\rm H}/J$. The dotted line indicates the string order parameter of the $S=1$ Heisenberg chain.
		(b) String order parameter $|\mathcal{O}_{\rm str}(-\infty,\infty)|$ as a function of $J_{\rm H}/J$. Inset: $|\mathcal{O}_{\rm str}(-\infty,\infty)|$ plotted against $\sqrt{J_{\rm H}/J}$ in the small $J_{\rm H}/J$ regime; the dashed line shows a fit near $J_{\rm H}/J=0$.
	}
	\label{fig:Heis}
\end{figure}

{\it Haldane Phase in the Heisenberg Ladder.}---
We first investigate how the Heisenberg ladder converges to the effective spin-1 Heisenberg chain [Eq.~\eqref{eq:s1hcain}] as $J_{\rm H}/J$ increases. Figure~\ref{fig:Heis}(a) shows the string correlation function $|\mathcal{O}_{\rm str}(i,j)|$ versus distance $|i-j|$ for several values of $J_{\rm H}/J$, computed on a $193 \times 2$ ladder. For all $J_{\rm H}/J$ values considered, the correlation function clearly saturates at large distances, confirming robust long-range string order.

In Fig.~\ref{fig:Heis}(b), we plot the extracted string order parameter as a function of $J_{\rm H}/J$. As $J_{\rm H}/J$ grows, the string order parameter continuously approaches the known spin-1 Heisenberg chain value, demonstrating that the ladder progressively realizes the effective spin-1 limit. As shown in the inset, the string order parameter follows the scaling form $|\mathcal{O}_\mathrm{str}(i,j)| \propto \sqrt{J_{\rm H}/J}$ in the small-$J_{\rm H}/J$ regime, indicating the immediate onset of the Haldane phase for any finite $J_{\rm H}>0$. Consistent with this, the spin gap opens simultaneously with the emergence of string order, as discussed in detail below.

\begin{figure}[tb]
	\centering
	\includegraphics[width=0.85\linewidth]{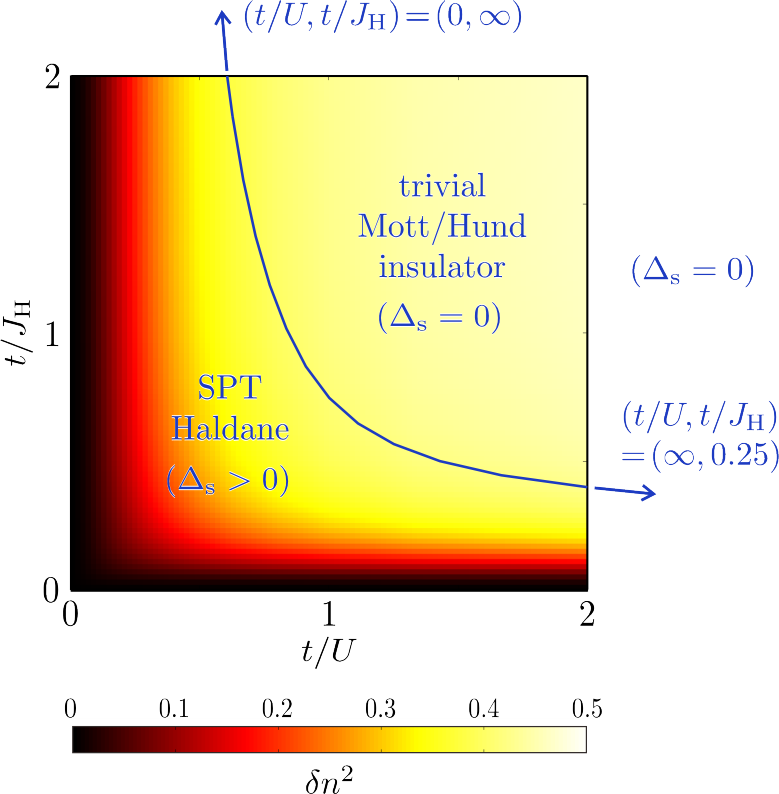}
	\caption{
		Charge fluctuations in the Hubbard ladder. Local charge variance $\delta n^2$ plotted as a function of $t/U$ and $t/J_H$. The solid line indicates the phase boundary separating the trivial Mott insulating phase from the topological Haldane phase.		
	}
	\label{fig:variance}
\end{figure}

{\it Charge Fluctuations in the Hubbard Ladder}---
We now turn to the Hubbard ladder [Eq.~\eqref{eq:HamHub}], which differs from the Heisenberg ladder by allowing double occupancy of lattice sites, thereby introducing charge fluctuations. To quantify these fluctuations, we examine the local charge variance, defined as $\delta n^2=\langle n_{\gamma,i}^2 \rangle-\langle n_{\gamma,i} \rangle^2$. At half filling, this quantity directly relates to double occupancy: $\delta n^2=2\langle n_{\gamma,i,\uparrow}n_{\gamma,i,\downarrow}\rangle$.

Figure~\ref{fig:variance} shows $\delta n^2$ as a function of $t/U$ and $t/J_{\rm H}$. Increasing $U$ naturally reduces double occupancy, suppressing $\delta n^2$ and pushing the system toward the spin-only limit. Increasing $J_{\rm H}$ similarly suppresses charge fluctuations, but through a distinct mechanism: FM Hund’s coupling energetically favors rung spin-triplet formation, thereby stabilizing well-defined spin-$\frac{1}{2}$ states and disfavoring doublon-holon pairs. Consequently, $\delta n^2$ decreases systematically. Thus, Coulomb repulsion $U$ and Hund’s coupling $J_{\rm H}$ cooperatively reduce charge fluctuations, stabilizing an effective low-energy spin description.

\begin{figure}[tb]
	\centering
	\includegraphics[width=1.0\linewidth]{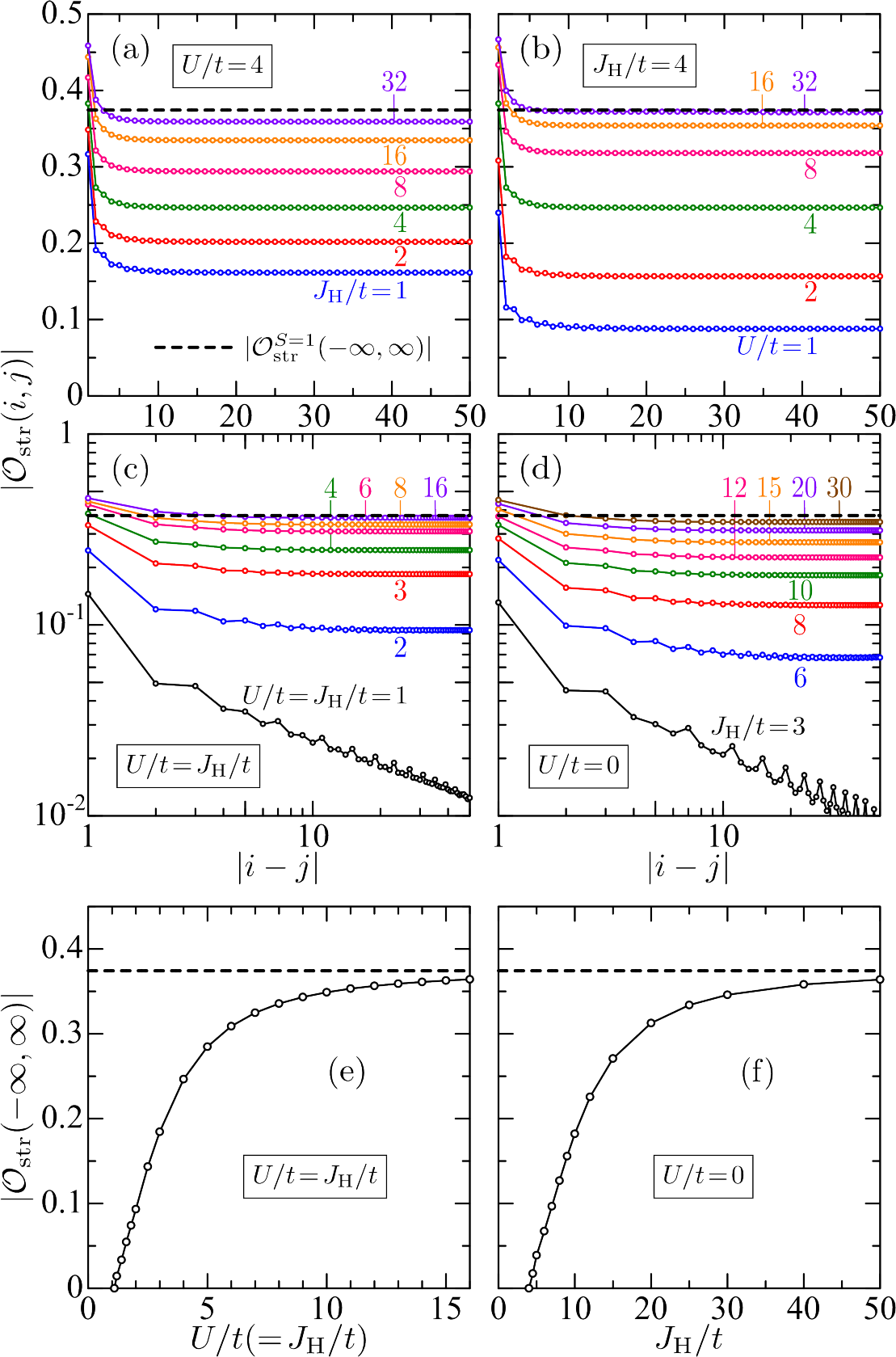}
	\caption{
		Topological string order in the Hubbard ladder.
		String correlation function $O_{\text{str}}(i,j)$ versus $|i-j|$ for (a) various $J_H/t$ at fixed $U/t=4$ and (b) various $U/t$ at fixed $J_H/t = 4$.
		Log-log plots of $O_{\text{str}}(i,j)$ versus $|i-j|$ for (c) various interaction strengths with $U/t = J_H/t$, and (d) various $J_H/t$ at fixed $U/t=0$.
		String order parameter as a function of (e) $U/t=J_H/t$ and (f) $J_H/t$ at fixed $U/t=0$. Dotted lines mark the spin-1 Heisenberg chain value.
	}
	\label{fig:Hub}
\end{figure}

{\it Topological Order in the Hubbard Ladder}.---
We now investigate how charge fluctuations influence topological order in the Hubbard ladder by examining the string correlation function $\mathcal{O}_{\rm str}(i,j)$. Figure~\ref{fig:Hub}(a) presents $|\mathcal{O}_{\rm str}(i,j)|$ as a function of distance $|i-j|$ for several values of $J_{\rm H}/t$ at fixed $U/t=4$, while Fig.~\ref{fig:Hub}(b) shows similar results for various $U/t$ at fixed $J_{\rm H}/t=4$, computed on a $97 \times 2$ ladder. In both cases, the string correlation saturates at finite values at large distances, confirming robust long-range order. With increasing $U/t$ or $J_{\rm H}/t$, charge fluctuations are progressively suppressed, driving the system closer to the effective spin-1 regime. Correspondingly, the saturated value, i.e., the string order parameter, approaches the spin-1 Heisenberg chain value. Thus, the Hubbard ladder can realize a stable Haldane phase despite moderate charge fluctuations.

However, when both $U/t$ and $J_{\rm H}/t$ are small and charge fluctuations dominate, the string correlations vanish at large distances, signaling the absence of long-range string order. Figure~\ref{fig:Hub}(c) shows a log-log plot of $|\mathcal{O}_{\rm str}(i,j)|$ versus distance for equal interaction strengths $U/t=J_{\rm H}/t$. At weak coupling ($U/t=J_{\rm H}/t=1$), a power-law decay is observed, indicating no long-range order. At stronger coupling ($U/t=J_{\rm H}/t=2$), however, finite long-range string order emerges, signaling a topological phase transition between these two interaction strengths. Notably, near the critical point, the string order parameter grows nearly linearly, unlike the square-root behavior observed in the Heisenberg ladder, before gradually approaching the spin-1 Heisenberg chain value at stronger interactions.

Long-range string order appears even at $U/t=0$, driven solely by Hund’s coupling. Figure~\ref{fig:Hub}(d) shows log-log plots of the string correlation for various $J_{\rm H}/t$ values at zero on-site repulsion. A clear transition from power-law decay ($J_{\rm H}/t=3$) to finite long-range order ($J_{\rm H}/t\geq 6$) is observed, reflecting strong suppression of charge fluctuations (double occupancy) by Hund’s coupling alone. This insulating phase stabilized purely by Hund’s coupling can thus be viewed as a novel ``Hund insulator''. The string order parameter, shown in Fig.~\ref{fig:Hub}(f), grows around $J_{\rm H}/t \approx 4.0$, approaching the spin-1 Heisenberg value at large $J_{\rm H}/t$, again following an approximately linear onset.

The resulting phase diagram, summarized in Fig.~\ref{fig:variance}, reveals a clear boundary between the trivial Mott/Hund insulating phase and the SPT Haldane phase, identified explicitly by the string order parameter. Notably, the Haldane phase remains stable under substantial charge fluctuations, persisting up to a local charge variance of approximately $\delta n^2 = 2\langle n_{\gamma,i,\uparrow} n_{\gamma,i,\downarrow}\rangle \lesssim 0.4$. Furthermore, the phase boundary approaches $J_{\rm H}=0$ in the $U = \infty$ limit, indicating a smooth crossover from the Hubbard to the Heisenberg ladder in the strongly correlated regime. Although our model setup is different, the qualitative agreement with the topological transition line reported in Ref.~\cite{Jazdzewska2023} supports the broader conclusion that the Haldane phase is generally stable against moderate charge fluctuations.

\begin{figure}[tb]
	\centering
	\includegraphics[width=1.0\linewidth]{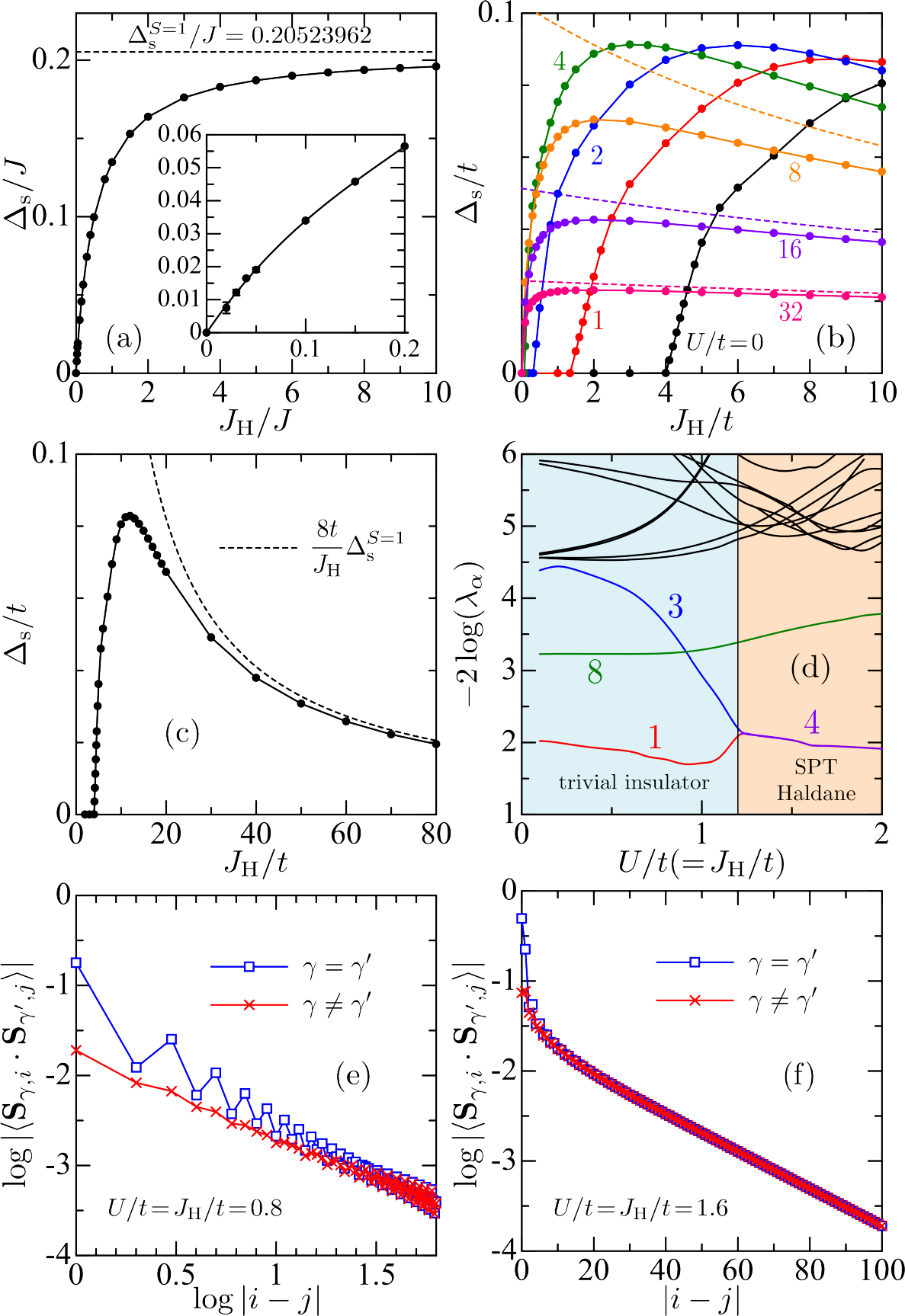}
	\caption{
		(a) Spin gap $\Delta_{\rm s}/J$ for the Heisenberg ladder as a function of $J_{\rm H}/J$. The dashed line marks the spin gap of the spin-1 Heisenberg chain [Eq.~\eqref{eq:s1hcain}]. Inset: Enlarged view of the small-$J_{\mathrm{H}}/J$ region.
		(b) $\Delta_{\rm s}/t$ for the Hubbard ladder as a function of $J_{\rm H}/t$ for several values of $U/t$.
		(c) $\Delta_{\rm s}/t$ as a function of $J_{\rm H}/t$ at $U=0$. Dashed lines in (b) and (c) indicate strong-coupling estimates.
		(d) Entanglement spectrum across the topological transition at $U/t=J_{\rm H}/t \approx 1.2$. Numbers beside the lines denote entanglement level degeneracies.
		(e,f) Spin–spin correlation function $\langle \mathbf{S}_{\gamma,i} \cdot \mathbf{S}_{\gamma',j} \rangle$ as a function of distance $r$, shown on (e) log-log and (f) semi-log scales for $U/t=J_{\rm H}/t=0.8$ and 1.6, respectively.
	}
	\label{fig:spingap}
\end{figure}

{\it Spin Gap and Edge states.}---
A finite spin gap is a defining feature of the Haldane phase. To confirm its presence, we compute the spin gap $\Delta_{\rm s}$, first considering the Heisenberg ladder. Figure~\ref{fig:spingap}(a) shows the spin gap $\Delta_{\rm s}/J$ versus $J_{\rm H}/J$. Consistent with the immediate onset of string order, a spin gap emerges immediately upon introducing finite $J_{\rm H}$, confirming that the Haldane phase appears throughout the entire $J_{\rm H}>0$ regime. At small $J_{\rm H}/J$, the gap grows approximately linearly, contrasting the singular, square-root onset of the string order. In the large-$J_{\rm H}/J$ limit, the gap approaches the known spin-1 Heisenberg chain value $\Delta_{\rm s}^{S=1}/J=0.20523962$.

The relationship between spin gap and string order in the Hubbard ladder is more subtle. Figure~\ref{fig:spingap}(b) plots $\Delta_{\rm s}/t$ as a function of $J_{\rm H}/t$ for various $U/t$. A finite spin gap exists broadly, disappearing only at small $U/t$ and $J_{\rm H}/t$. A strong-coupling analysis yields an effective exchange $J_{\rm eff}=4t^2/(U+J_{\rm H}/2)$, predicting the asymptotic behavior $\Delta_{\rm s}/t=\{4t/(U+J_{\rm H}/2)\}\Delta_{\rm s}^{S=1}$. Also, a finite gap persists even at $U/t=0$ for $J_{\rm H}/t\gtrsim4$, reflecting an effective repulsion generated purely by strong Hund’s coupling [Fig.~\ref{fig:spingap}(c)]. Here, the gap asymptotically follows $\Delta_{\rm s}/t=(8t/J_{\rm H})\Delta_{\rm s}^{S=1}$. Details of the strong-coupling analysis are given in the Supplemental Material~\cite{SM}.

Overlaying the finite spin-gap region onto Fig.~\ref{fig:variance}, we find a perfect match with the finite string-order region. This nontrivial correspondence strongly supports the identification of this regime as a genuine Haldane phase. We also find that slightly away from half filling both the nonlocal string order and the spin gap remain finite in the infinitesimal doping limit (see the Supplemental Material~\cite{SM}).

To further confirm the topological character, we analyze the ES. Figure~\ref{fig:spingap}(d) shows the ES across the topological transition at $U/t=J_{\rm H}/t\approx1.2$. The lowest entanglement level is non-degenerate in the trivial insulating phase, becoming fourfold degenerate immediately upon entering the Haldane phase, consistent with the expected SPT order.

Finally, we examine real-space spin correlations to further distinguish the gapped Haldane phase from gapless trivial states. Figures~\ref{fig:spingap}(e,f) present intra- and inter-leg spin correlations $\langle \mathbf{S}_{\gamma,i}\cdot\mathbf{S}_{\gamma',j}\rangle$ as functions of distance $|i-j|$. In the Haldane phase ($U/t=J_{\rm H}/t=1.6$), the correlations decay exponentially as $\langle \mathbf{S}_{\gamma,i} \cdot \mathbf{S}_{\gamma^\prime,j} \rangle \sim \exp(-|i-j|/\xi)$ ($\xi>0$), whereas in the gapless regime ($U/t=J_{\rm H}/t=0.8$), correlations show power-law decay, signaling criticality. Notably, within the Haldane phase, intra- and inter-leg correlations rapidly coincide, validating the emergence of effective spin-1 degrees of freedom and reinforcing the interpretation of the ladder as an effective spin-1 chain at low energies.

{\it Summary.}---
We have investigated the stability of the Haldane phase in Heisenberg and Hubbard ladders with interchain Hund’s coupling. While the Heisenberg ladder realizes a spin-1 Haldane phase for any finite $J_{\rm H} > 0$, the inclusion of charge fluctuations in the Hubbard ladder leads to qualitatively distinct behavior. Using DMRG simulations, we demonstrated that the Haldane phase—characterized by long-range string order, a finite spin gap, and degeneracies in the ES—persists over a broad parameter range in the Hubbard ladder, but ultimately breaks down when both $U$ and $J_{\rm H}$ are small.

These results establish that the Haldane phase remains robust against moderate charge fluctuations, but is destabilized when electronic itinerancy becomes too strong. Our findings highlight the subtle interplay between spin and charge degrees of freedom in realizing SPT phases, and delineate the conditions under which topological order can persist in correlated itinerant systems.

{\it Acknowledgments.}---
We thank Ulrike Nitzsche for technical assistance. This project is funded by the German Research Foundation (DFG) via the projects A05 of the Collaborative Research Center SFB 1143 (project-id 247310070). Part of the DMRG calculations in this work were performed using the ITensor Software Library~\cite{itensor}.

\vspace*{5.0mm}
\noindent
The data used to generate the figures in this manuscript are available in Ref.~\cite{data}.

\bibliography{halffilledHaldane}% Produces the bibliography via BibTeX.

\end{document}

%\widetext

\begin{center}
	\large
	{\bf Supplemental Material for\\ ``
		Instability of the Haldane Phase: Roles of Charge Fluctuations and Hund's Coupling''}
	
	\normalsize
	\vspace{3.0mm}
	Satoshi Nishimoto
	
	\small
	
	\vspace{1.5mm}
	{\it
		Department of Physics, Technical University Dresden, 01069 Dresden, Germany\\
		Institute for Theoretical Solid State Physics, IFW Dresden, 01069 Dresden, Germany
	}
\end{center}

\begin{abstract}
\end{abstract}
%\date{\today}
\maketitle

\vspace{-1em}
\tableofcontents
\vspace{1em}
\hrule
\vspace{1.5em}

\newpage

\subsection{I. Determination of the string order parameter and finite-size scaling}

\begin{figure}[h]
	\centering
	\includegraphics[width=0.6\linewidth]{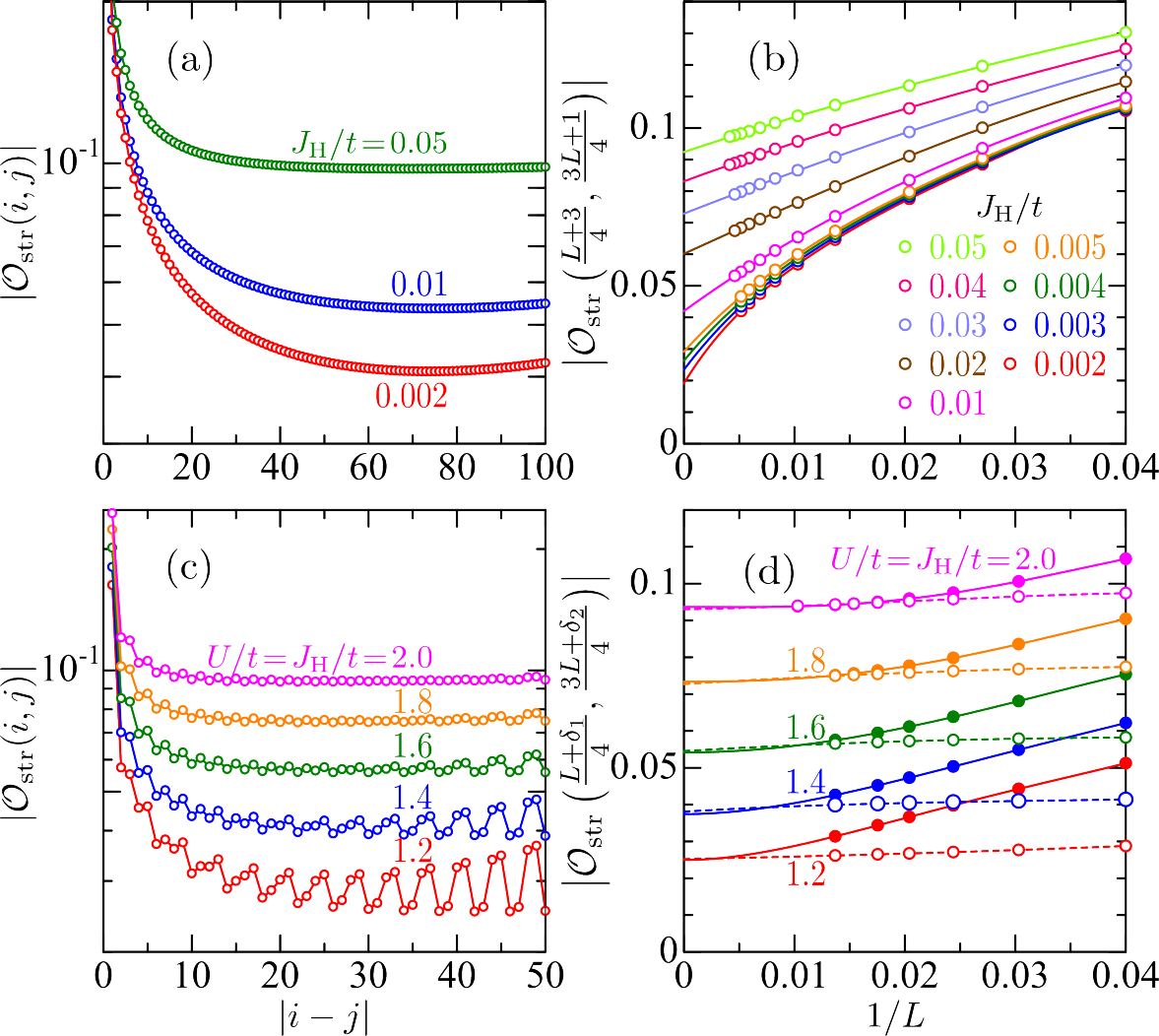}
	\caption{
	String correlations and finite-size scaling in the Heisenberg and Hubbard ladders.
	(a,b) Heisenberg ladder: (a) $\mathcal{O}_\mathrm{str}(i,j)$ vs distance $|i-j|$ for several $J_{\rm H}/J$ on a $193\times2$ ladder; (b) finite-size scaling of the string order parameter $|\mathcal{O}_\mathrm{str}((L+3)/4,(3L+1)/4)|$.
	(c,d) Hubbard ladder: (c) $|\mathcal{O}_\mathrm{str}(i,j)|$ vs $|i-j|$ for several $U/t$ (with $J_{\rm H}/t=U/t$) on a $73\times2$ ladder; (d) corresponding finite-size scaling as in (b). Solid symbols denote  $|\mathcal{O}_\mathrm{str}(\frac{L+3}{4},\frac{3L+1}{4})|$ and open symbols $|\mathcal{O}_\mathrm{str}(\frac{L+3}{4},\frac{3L-3}{4})|$.
	All data use OBC; as in the main text, one spin-$\tfrac12$ site is removed at each edge (first and $L$th rungs) to suppress edge-mode artifacts, and correlators are evaluated about the ladder center.
	}
	\label{fig:scaling_SM}
\end{figure}

As stated in the main text, the string order parameter is defined as the long-distance value of the string correlation function $\mathcal{O}_{\mathrm{str}}(i,j)$. When the order is \emph{robust}, the correlation converges rapidly with distance and finite-size effects from open boundaries (Friedel oscillations) are weak; in that case one can read off the order parameter directly from sufficiently long ladders (see main-text Figs. 2(a) and 4(a-d)). By contrast, when the order is \emph{not robust}, the convergence of $|\mathcal{O}_\mathrm{str}(i,j)|$ with $|i-j|$ is slow and the correlations are visibly distorted by edge effects, making a single-size estimate unreliable. We therefore employ finite-size scaling. In all analyses of this section we follow the main-text setup and remove one spin-$\tfrac12$ site at each edge (first and $L$th rungs; see Fig.~1 in the main text) to quench edge-mode contributions and to center the string segment.

\emph{Heisenberg ladder}. For $J_{\rm H}/J \gtrsim 0.1$ the long-distance plateau is clear in finite systems, whereas for $J_{\rm H}/J \lesssim 0.1$ the convergence cannot be unambiguously confirmed on ladders with $L\sim100$-$200$ rungs. Figure~\ref{fig:scaling_SM}(a) shows $\mathcal{O}_{\mathrm{str}}(i,j)$ versus $|i-j|$ on a $193\times2$ ladder for several small $J_{\rm H}/J$: after an apparent approach to a plateau, the curve grows again at larger distances due to edge effects, obscuring the thermodynamic limit. We therefore determine the order parameter by finite-size scaling: on $L\times2$ ladders we evaluate $\left|\mathcal{O}_{\mathrm{str}}\!\left(\tfrac{L+3}{4},\tfrac{3L+1}{4}\right)\right|$ for odd $L=25,\dots,241$, and extrapolate it as a function of $1/L$, as shown in Fig.~\ref{fig:scaling_SM}(b). We use third- or higher-order polynomials in $1/L$; the extrapolated values are plotted in main-text Fig. 2(b).

\emph{Hubbard ladder}. Figure~\ref{fig:scaling_SM}(c) shows $|\mathcal{O}_{\mathrm{str}}(i,j)|$ versus $|i-j|$ on a $73\times2$ ladder for $U/t=J_{\rm H}/t=1.2,\,1.4,\,1.6,\,1.8$, and $2.0$. Since the topological transition lies near $U/t=J_{\rm H}/t\simeq1$, the string order is expected to be less robust in this range. The overall behavior resembles the Heisenberg case but with more pronounced oscillations, which obscure a direct readout. Accordingly, we scale two central estimators at separations $|i-j|=L/2$ and $L/2-1$, $\Big|\mathcal{O}_{\mathrm{str}}\!\Big(\tfrac{L+3}{4},\tfrac{3L+1}{4}\Big)\Big|$, and $\Big|\mathcal{O}_{\mathrm{str}}\!\Big(\tfrac{L+3}{4},\tfrac{3L-3}{4}\Big)\Big|$,and fit each sequence versus $1/L$ with $f(1/L)=a_0+a_2/L^2+a_3/L^3+a_4/L^4$. As shown in Fig.~\ref{fig:scaling_SM}(d), both sequences extrapolate to the same thermodynamic value $a_0$ within errors, corroborating the finite-size-scaling procedure.

\newpage

\section{I\hspace{-1.2pt}I. Possible topological order away from half filling}

\begin{figure}[h]
	\centering
	\includegraphics[width=0.6\linewidth]{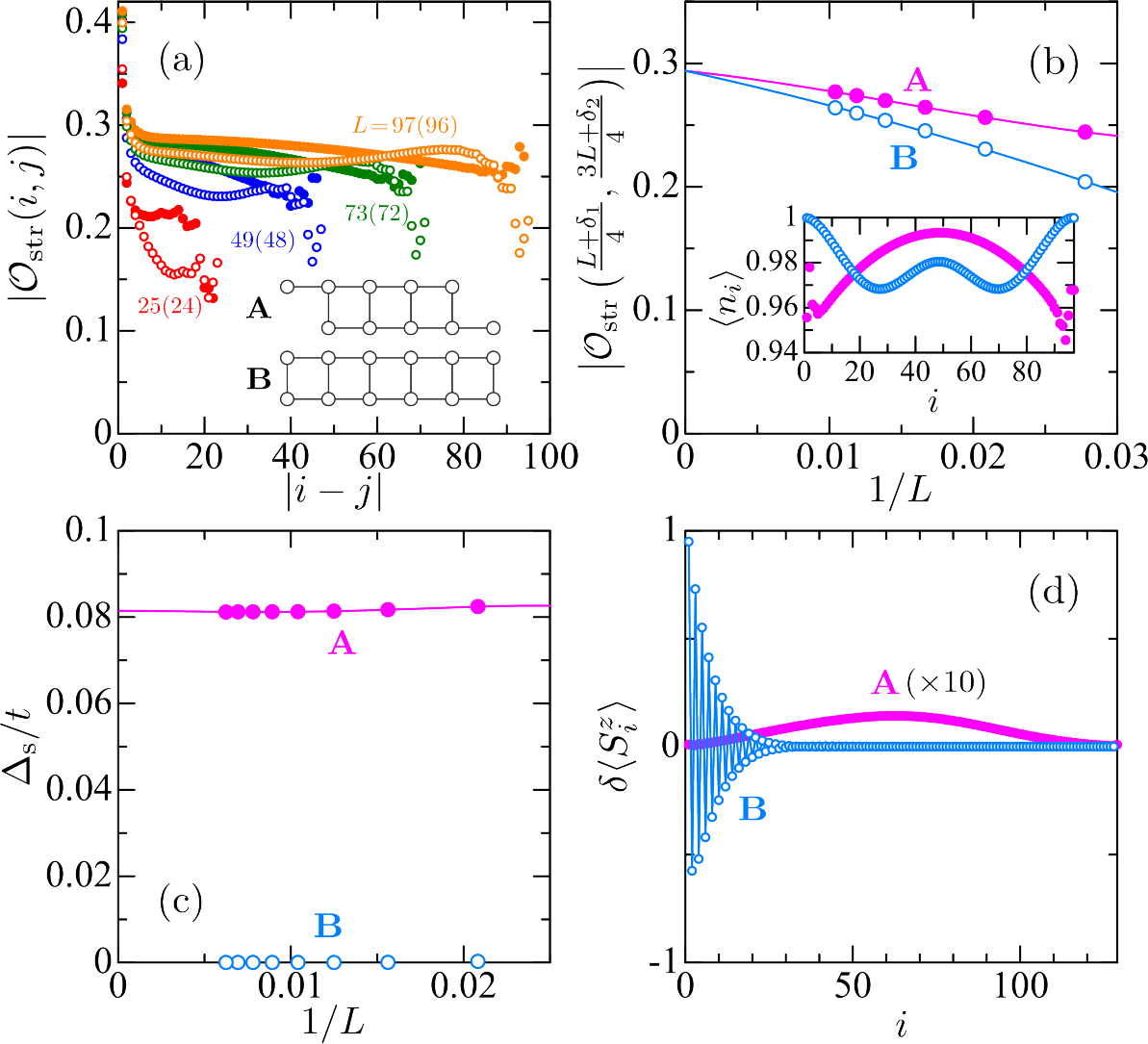}
	\caption{
		String order and spin gap under infinitesimal hole doping.
		DMRG data for $U/t=4.0$ and $J_{\rm H}/t=8.0$ on $L\times2$ ladders with \emph{four holes}
		(so $n=(2L-4)/(2L)\to1^-$ as $L\to\infty$).
		(a) $|\mathcal{O}_\mathrm{str}(i,j)|$ versus distance $|i-j|$ for several $L$.
		Solid (open) symbols correspond to \textbf{type-A} (\textbf{type-B}) clusters; see text.
		(b) Finite-size scaling of the central estimator
		$|\mathcal{O}_\mathrm{str}(\frac{L+\delta_1}{4},\frac{3L+\delta_2}{4})|$ with $(\delta_1,\delta_2)=(3,1)$ for type-A
		and $(0,0)$ for type-B (so that $|i-j|=L/2$ in both cases).
		Inset: charge density $\langle n_i\rangle$ for $L=97$ (A) and $96$ (B), showing a smooth density without phase separation.
		(c) Finite-size scaling of the spin gap $\Delta_{\rm s}/t$ for type-A and type-B clusters.
		(d) Spinon distribution $\delta\langle S^z_i\rangle\equiv \langle S^z_i\rangle_{S^z_{\rm tot}=1}-\langle S^z_i\rangle_{S^z_{\rm tot}=0}$
		in the first excited triplet for $L=129$ (A) and $128$ (B).
	}
	\label{fig:doping_SM}
\end{figure}

We examine whether the Haldane phase persists \emph{slightly} away from half filling by studying ladders doped with a fixed number of holes ($N_h=4$). As $L$ increases, the hole density $\delta=N_h/(2L)$ scales to zero, so this probes the \emph{infinitesimal-doping} limit $n\to1^-$.

Figure~\ref{fig:doping_SM}(a) shows the string correlation $|\mathcal{O}_\mathrm{str}(i,j)|$ versus distance $|i-j|$ for several $L$.
We consider two open-boundary cluster types. \textbf{Type-A} uses odd $L$ and the central evaluation points
$(i,j)=\big(\tfrac{L+3}{4},\tfrac{3L+1}{4}\big)$, which, upon mapping to an $S=1$ chain,
effectively removes the edge-mode contribution (intuitively, the end $S\!=\!1/2$ modes are quenched).
\textbf{Type-B} uses even $L$ and $(i,j)=\big(\tfrac{L}{4},\tfrac{3L}{4}\big)$, for which edge modes are present in the Haldane
regime. While the correlation increases with $L$, its $L\!\to\!\infty$ limit is not obvious from a single size,
so we perform a finite-size scaling of the central estimator taken at $|i-j|=L/2$. As shown in Fig.~\ref{fig:doping_SM}(b), both sequences extrapolate to $|\mathcal{O}_\mathrm{str}(-\infty,\infty)| \approx 0.294$, very close to the half-filled value $0.2939$ (same parameters), indicating that half filling is \emph{not} a singular point for the string order in the $n\to1^-$ limit. The inset of Fig.~\ref{fig:doping_SM}(b) shows
$\langle n_i\rangle$ for representative sizes: the density is less than 1 near the center and does not exhibit phase separation, confirming that the system is indeed doped away from half filling. This means that the system is metallic.

Consistent with a finite string order, the spin sector remains gapped in type-A clusters: Fig.~\ref{fig:doping_SM}(c) shows that $\Delta_{\rm s}(L)$ extrapolates to $\Delta_{\rm s}/t \approx 0.081$, in excellent agreement with the half-filled value $0.079640$. In contrast, type-B clusters extrapolate to (nearly) zero spin gap due to the presence of edge modes. This interpretation is corroborated by the spinon distribution $\delta\langle S^z_i\rangle$ in the first triplet (Fig.~\ref{fig:doping_SM}(d)): for type-A the flipped spin is localized in the bulk, whereas for type-B it resides near the edges, signaling (almost) free end spins.

In summary, for the parameters $U/t=4.0$ and $J_{\rm H}/t=8.0$ the Haldane string order and the spin gap \emph{survive in the infinitesimal-doping limit} $n\to1^-$. A systematic study at finite doping is left for future work.

\newpage

\section{I\hspace{-1.2pt}I\hspace{-1.2pt}I. Strong-coupling expansion of the Hubbard ladder}

The Hamiltonian of our Hubbard ladder is given by
\begin{align}
	\nonumber
	H_{\mathrm{Hub}} = -t \sum_{\gamma=1,2} \sum_{i,\sigma} \left( c^\dagger_{\gamma,i,\sigma} c_{\gamma,i+1,\sigma} + \text{H.c.} \right)\\
	+ U \sum_{\gamma,i} n_{\gamma,i,\uparrow} n_{\gamma,i,\downarrow}- J_{\rm H} \sum_i \mathbf{S}_{1,i} \cdot \mathbf{S}_{2,i},
	\label{eq:HamHub}
\end{align}
with intraleg hopping $t$, on-site repulsion $U$, and ferromagnetic (FM) Hund coupling $J_{\rm H}\!>\!0$ on each rung.

At half filling and $U,J_{\rm H}\!\gg\!t$, the low-energy subspace has one electron per site. On a rung,
$-J_{\rm H}\mathbf{S}_{1,i}\!\cdot\!\mathbf{S}_{2,i}$ favors the triplet
($E_{T}=-J_{\rm H}/4$) over the singlet ($E_{S}=+3J_{\rm H}/4$), with a gap $J_{\rm H}$.
A virtual hop along a leg creates one doubly occupied site (cost $U$) and breaks two neighboring
rung triplets (loss $J_{\rm H}/4$ on each), so the excitation energy is
\begin{equation}
	\Delta \;=\; U + \frac{J_{\rm H}}{2}.
	\label{eq:DeltaS}
\end{equation}
A Schrieffer-Wolff transformation to $\mathcal{O}(t^2/\Delta)$ generates an antiferromagnetic superexchange
along each leg with
\begin{equation}
	J_{\rm leg} \;=\; \frac{4t^2}{U+J_{\rm H}/2},
	\label{eq:JlegS}
\end{equation}
yielding the effective spin-$\tfrac12$ ladder with FM rungs,
\begin{equation}
	H_{\rm Heis} \;=\;
	J_{\rm leg}\sum_{\gamma=1,2}\sum_i \mathbf{S}_{\gamma,i}\!\cdot\!\mathbf{S}_{\gamma,i+1}
	\;-\; J_{\rm H}\sum_i \mathbf{S}_{1,i}\!\cdot\!\mathbf{S}_{2,i}.
	\label{eq:HeisLadderS}
\end{equation}

Our mechanism realizes an emergent $S=1$ per rung from two $S=\tfrac12$ aligned by ferromagnetic Hund's coupling. This is distinct from intrinsic $S=1$ ions (e.g., octahedral Ni$^{2+}$), yet at low energy both are adiabatically connected to the same Haldane symmetry-protected topological phase. The rung composite leaves small biquadratic corrections and characteristic high-energy rung-singlet excitations, and it controls the response to charge fluctuations and doping. In the regime $J_{\rm H}\!\gg\!J_{\rm leg}$, projecting onto the rung-triplet manifold produces an effective $S=1$ bilinear-biquadratic chain,
\[
H_{\rm eff}=J_1\sum_i \mathbf T_i\!\cdot\!\mathbf T_{i+1}
+K\sum_i(\mathbf T_i\!\cdot\!\mathbf T_{i+1})^2,\qquad
J_1\simeq \frac{J_{\rm leg}}{2},\quad
K=\mathcal O\!\Big(\frac{J_{\rm leg}^2}{J_{\rm H}}\Big),
\]
which is what underlies Fig. 5. In our parameter window $|K/J_1|\ll1$; including $K$ only refines the quantitative agreement. The derivation and coefficients, as well as an optional two-rung matching procedure to extract $(J_1,K)$ non-perturbatively.